\documentclass[proceedings]{JHEP3}
\PrHEP{ hep2001}

\usepackage{epsfig,multicol}   

\newbox\mybox
\newcommand\fverb{\setbox\mybox=\hbox\bgroup\verb}
\newcommand\fverbdo{\egroup\medskip\noindent\fbox{\unhbox\mybox}\ }
\newcommand\fverbit{\egroup\item[\fbox{\unhbox\mybox}]}

\title{\bf The role of jet quenching in the $\bar{p} \geq \pi^-$ anomaly at RHIC}

\author{ \bf \speaker{Ivan Vitev}, Miklos Gyulassy,  and 
               Peter L\'evai\thanks{KFKI Research Institute for Particle and
               Nuclear Physics, P.O.Box 49, Budapest 1525, Hungary.} 
\ \ \ \thanks{This work is supported by U.S.DOE grant DE-AC0376SF00098 
and Hungarian grant OTKA-T032796.}\\
        Columbia University, 538 West 120-th Street, New York, NY 10027, USA\\
        E-mail: \email{ivitev@nt3.phys.columbia.edu},
        \email{gyulassy@nt3.phys.columbia.edu}, \email{plevai@rmki.kfki.hu} }

\conference{EPS-HEP 2001}

\abstract{ Preliminary PHENIX data on $Au+Au$ reactions at $\sqrt{s}= 130$ AGeV suggest
that $\bar{p}$ yields may exceed $\pi^-$ at high $p_{\rm T} >
   2$~GeV/c. We propose that jet quenching in central collisions suppresses the
   hard  PQCD component of the spectra in central $A+A$ reactions,
   thereby exposing a novel non-perturbative component of baryon dynamics. We
   suggest that baryon junctions provide a possible explanation of
   the anomalous component. We predict that the $\bar{p} \geq \pi^-$ anomaly is
   limited to a  finite $p_{\rm T}$ window and decreases with
   increasing impact parameter.   }

\begin{document}

\section{Introduction}

Preliminary  RHIC data on $\sqrt{s}=130$~AGeV $Au+Au$ reactions
have revealed a number of qualitatively 
new phenomena at moderate high $p_T \sim 2-6$ GeV/c.
\begin{enumerate}
\item The  high $p_{\rm T}$ spectra of $\pi^0$ in the 10\% central
collisions were found by PHENIX 
to be suppressed by a factor 3-4 relative to PQCD predictions scaled by
nuclear geometry ($T_{AB}({\bf b})$, the number of binary collisions
at impact parameter ${\bf b}$). In contrast, the inclusive charged particle spectrum 
was reported to be suppressed only by a factor of 2~\cite{phen}.
\item Non-central collisions STAR
measured elliptic flow consistent with hydro predictions in the soft 
$p_{\rm T}$ region. However, this asymmetry saturates
at $v_2\sim 0.15$ above $p_T>2\;{\rm GeV/c}$~\cite{star,gvw}. 
\item  Even more surprisingly  
PHENIX reported~\cite{julia} that the high $p_{\rm T} > 2$ GeV
flavor composition of positive and negative  hadron yield 
may actually be dominated by anti-protons and protons respectively!
\end{enumerate}

In addition to the results listed above, there are already
global indications from preliminary STAR data~\cite{nuxuqm01}
of novel baryon production dynamics in $A+A$.
In central collisions the midrapdity $\bar{p}/p\sim 0.65$ and 
$dN^B/dy \simeq  dN^p/dy-dN^{\bar{p}}/dy \sim 15$.
Thus a very significant baryon number transport (stopping)
five units of rapidity from the fragmentation regions
and copious $B\bar{B}$ production seems
to have been observed. This is consistent with expectations based
on a non-perturbative  baryon-junction mechanism
for baryon number transport~\cite{junc}. 
In this paper we propose that  the baryon/meson anomaly  together with     
the  different degree of suppression of $\pi^0$ and inclusive charged hadrons   
in central collisions may be an unexpected new consequence of jet quenching 
which unveils  a novel  non-perturbative component of baryon production 
dynamics~\cite{gvjunc}.

Our  approach is to extend the two component hybrid model 
introduced in Ref.~\cite{gvw} considering an 
anomalously large contribution to $\bar{p}$ production at moderate $p_{\rm T}$
motivated by baryon junction dynamics~\cite{gvjunc}.  
The baryon junction picture implies a simple
relation between the mean inverse slopes of pions and protons, 
$T_p \simeq \sqrt{3} T_\pi$ (in the hassles limit), for the soft to moderate
$p_{\rm T}$ non-perturbative part of the spectrum.  
We compute jet quenching using the non-abelian energy loss from
the Gyulassy-Levai-Vitev  (GLV)  formalism~\cite{glv}.
Our analysis suggests  that  the enhancement of $\bar{p}/\pi^-$ ratio 
is limited to a finite 
$p_{\rm T}$ range 2-5~GeV and to central and semi-central collisions. 
Beyond this range,  the hadron ratios are 
predicted to converge to the jet quenched PQCD base.

\section{Reference flavor composition in PQCD}

The standard PQCD approach
expresses the differential hadron cross section in $p+p\rightarrow h+X$ 
as a convolution of the measured structure functions 
$f(x_\alpha,Q_\alpha^2)_{\alpha/p}$  for the interacting partons 
($\alpha = a,b$),  with the parameterized fragmentation function 
$D_{h/c}(z,Q^2_c)$ for
the leading rescattered parton $c$ into a hadron of flavor $h$ and the
elementary parton-parton cross sections 
$d\sigma^{(ab \rightarrow cd)}/d\hat{t}$.
The initial parton broadening  
can  be accounted for by a normalized  $k_{\rm T}$-smearing distribution. 
A simple functional form is the Gaussian smearing
$f(k_{\rm T}) = {e^{-k_{\rm T}^2/\langle k_{\rm T}^2 \rangle}}
/ {\pi\langle k_{\rm T}^2 \rangle }$ 
where $\langle k_{\rm T}^2 \rangle \simeq 0.6-1.0$~GeV$^2$ in $pp$ collisions. 
The invariant hadron inclusive cross section is then given by
\begin{eqnarray}
E_{h}\frac{d\sigma_h^{pp}}{d^3p} &=&
K   \sum_{abcd} \int\! dz_c dx_a 
dx_b \int d^2{\bf k}_{{\rm T}a} d^2{\bf k}_{{\rm T}b}  
\, f({\bf k}_{{\rm T}a})f({\bf k}_{{\rm T}b})
f_{a/p}(x_a,Q^2_a) f_{b/p}(x_b,Q^2_b) \nonumber \\[1ex]
&\;&  D_{h/c}(z_c,{Q}_c^2) 
 \frac{\hat{s}}{\pi z^2_c} \frac{d\sigma^{(ab\rightarrow cd)}}
{d{\hat t}} \delta(\hat{s}+\hat{u}+\hat{t}) \; ,
\label{hcrossec}
\end{eqnarray}
where $x_a, x_b$ are the initial momentum fractions carried 
by the interacting partons, $z_c=p_h/p_c$ is the momentum fraction carried 
by the observed hadron.

Fig.~1a shows comparison between the PQCD calculation and the negative hadron
multiplicities as measured by the UA1 experiment at 
$\sqrt{s}=200$~GeV. For our first fit we used 
the Gl\"{u}ck-Reya-Vogt structure functions (GRV 94) 
and Binnewies {\em et al.} (B) fragmentation functions. For our second 
fit we used the CETQ 5M structure functions and the
Kniehl {\em et al.} (KKP) fragmentation functions~\cite{funcs}.      
Eq.(\ref{hcrossec})
reproduces well the shape and norm of the data for 
$p_{\rm T} > 1$~GeV.

\begin{center}
\vspace*{7.cm}
\includegraphics{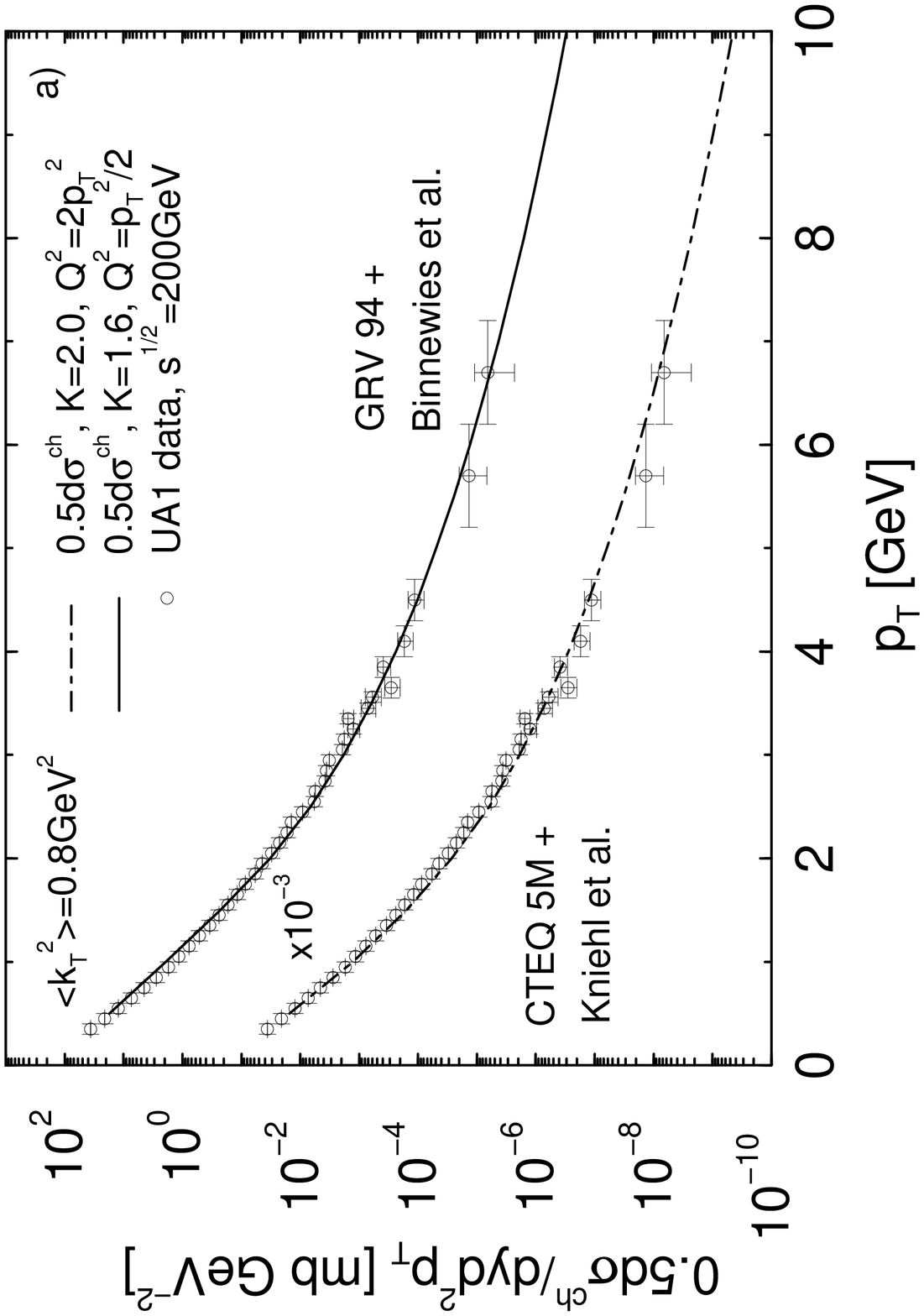}
\vspace{-2cm}
\includegraphics{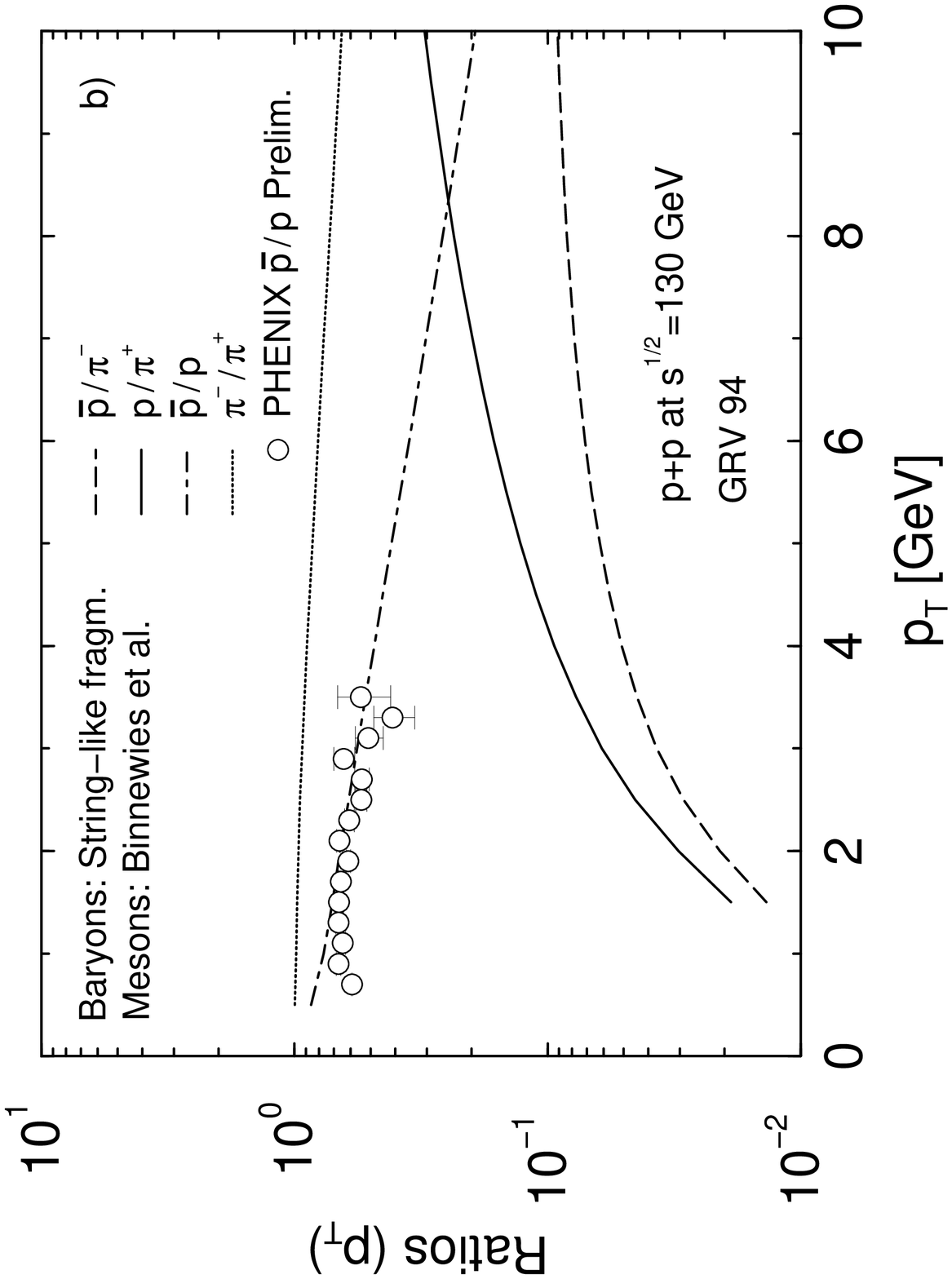}
\vspace{-2.9cm}

\vspace{2.cm}

\begin{minipage}[t]{14cm}
         {\small \bf Figure 1:} {\small {\bf a)}  
Negative hadron cross sections are shown
as a function of $p_{\rm T}$. Experimental UA1 data 
at $\sqrt{s}=200$~GeV
is compared to PQCD calculations with GRV94 pdf + B fragmentation
   functions ($Q^2= p_{\rm T}^2/2$) and 
CTEQ 5M pdf + KKP fragmentation function
   ($Q^2=2 p_{\rm T}^2$) where $\langle k_{\rm T}^2 \rangle = 0.8$~GeV$^2$ 
in both cases.
%compared to PQCD calculations with $\langle k_{\rm T}^2 
%\rangle=0.8$~GeV$^2$, $Q^2=p_T^2/2$ and $Q^2=2p_T^2$ for the GRV 94 + B
%and CTEQ 5M + KKP structure + fragmentation functions respectively.
{\bf b)} The PQCD predictions for the 
$p/\pi^+,\; \bar{p}/\pi^-, \; \pi^-/\pi^+$  and $\bar{p}/p$ ratios 
as a function of $p_{\rm T}$ are shown 
for $\sqrt{s}=130$~GeV.
%transverse momenta up to 
%10~GeV for $pp$ collisions at $\sqrt{s}=130$~GeV.
}
\end{minipage}
\end{center}
Fig.~1b, however, shows that although PQCD fragmentation
can fit the $\pi^-/\pi^+$  and even get quantitatively correct 
$\bar{p}/p$ ratios for $p_{\rm T} > 3-4$~GeV the  $p/\pi^+$ and  $\bar{p}/\pi^-$
differ from  the values suggested by data ($\sim 1$) by an order of magnitude. 
Even large uncertainties in the gluon fragmentation functions 
($\sim 3$)~\cite{gvjunc}  cannot account for such a discrepancy. 
The observed large (anti)proton 
enhancement in $Au+Au$ collisions in the region of $p_{\rm T}= 2-4$~GeV~\cite{julia}  
signals of interesting interplay between competing possibly novel 
physical effects. We propose that the mechanism which accounts for the copious 
production of baryons and anti-baryons might be related to baryon
junctions. Other possibilities include relativistic hydrodynamics and boosted
thermal sources.

\section{Medium effects and baryon junctions}

We first discuss the medium effects on the high $p_{\rm T}$ particle
spectra. The  medium induced gluon radiation spectrum is calculated as in~\cite{glv} and  
modifies  the fragmentation function  $D_{h/c}(z_c)$  in
Eq.(\ref{hcrossec}) as follows
\begin{equation}
%D^\prime_{h/c}(z_c) = \int_{x_{\rm min}}^1 dx\; \left[ \frac{1}{1-x \, n(x)} 
%D_{h/c} \left(\frac{z}{1-x\, n(x)} \right) +n(x) \frac{1}{x} 
%D_{h/g} \left(\frac{z}{x} \right) \;\right] \;\;,
D^\prime_{h/c}(z_c) = \int_{x_{\rm min}}^1 dx\; P(x)\left[ \frac{1}{1-x} 
D_{h/c} \left(\frac{z}{ 1-x } \right) +  \frac{1}{x} 
D_{h/g} \left(\frac{z}{x} \right) \;\right] \;\;,
\label{modfr}
\end{equation}  
where $x = \omega / E_{\rm jet}$, $P(x)$ is the gluon radiative probability, 
and the second term in Eq.(\ref{modfr}) accounts for the gluon feedback in the
system. Transverse expansion was shown not to affect the {\em azimuthally averaged}
quenching pattern~\cite{gvw2}  

The string and baryon junction description of the soft non-perturbative part
of the spectrum is parametized in terms of the mean inverse slopes of hadrons
$T_0^\alpha$ as in~\cite{gvw}. 
The baryon junction picture, however, 
suggest different $T_0^\pi$ and $T_0^p$. 
Large $T_0^{p}$ may arise from the small ($\delta r_\perp\sim
1/2 T_0^{\bar{p}} \sim 1/4$~fm) intrinsic (non-perturbative)
spatial structure of Junction-anti-Junction loops
and the smaller junction trajectory slope
$\alpha_J^\prime\approx \alpha_R^\prime/3$~\cite{junc}.
The latter implies that the effective string tension
is  three times higher than $1/(2\pi\alpha_R^\prime)\approx 1$ GeV/fm
leading in the massless limit to 
\begin{equation}
\langle p_{\rm T}^2 \rangle_J\approx 
3 \langle p_{\rm T}^2 \rangle_R \, ,  \qquad  T_0^p \simeq \sqrt{3} T_0^\pi \;\; .
\label{sloperelat}
\end{equation}

\begin{center}
\vspace*{7.cm}
\includegraphics{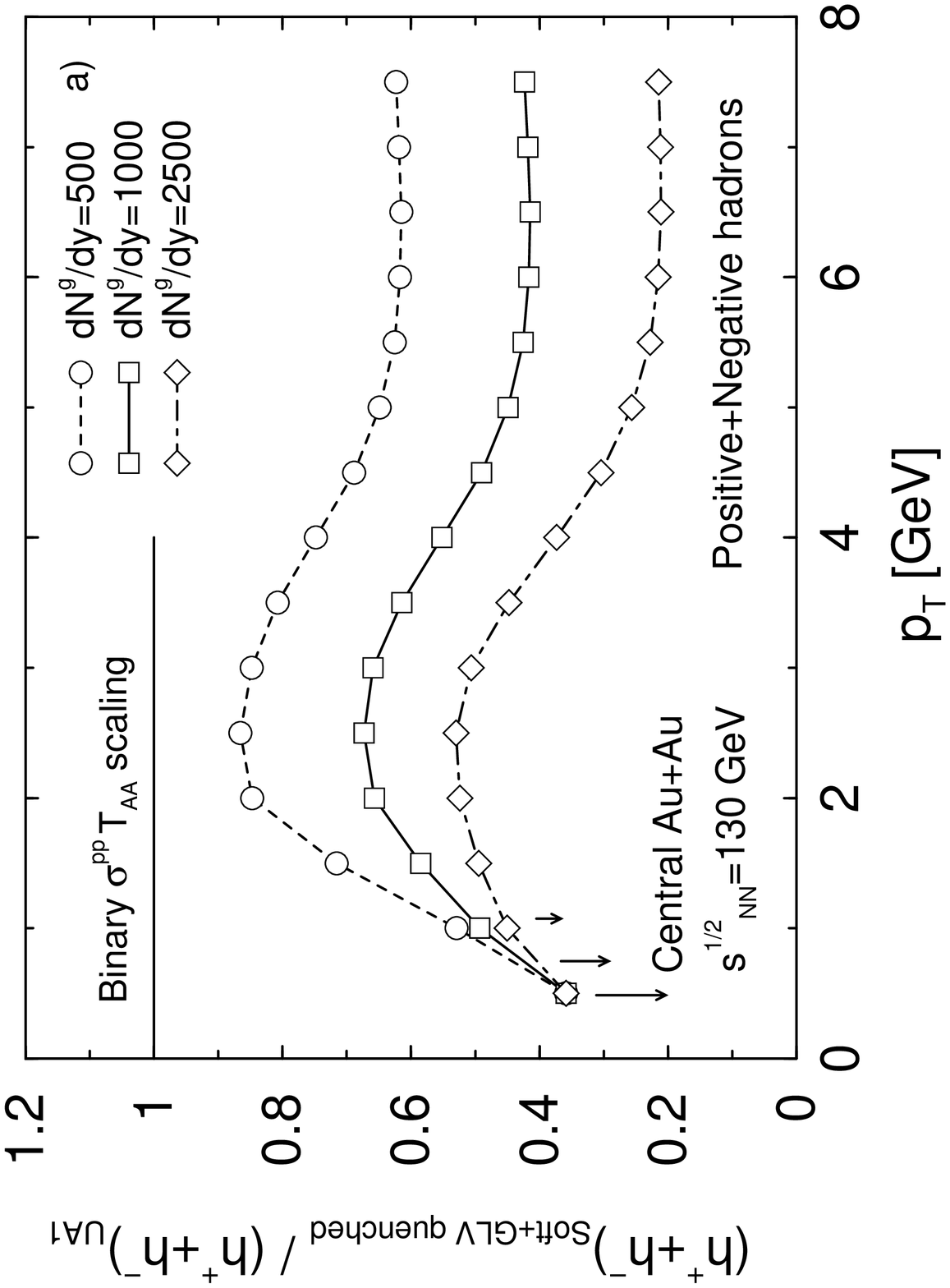}
\vspace{-2cm}
\includegraphics{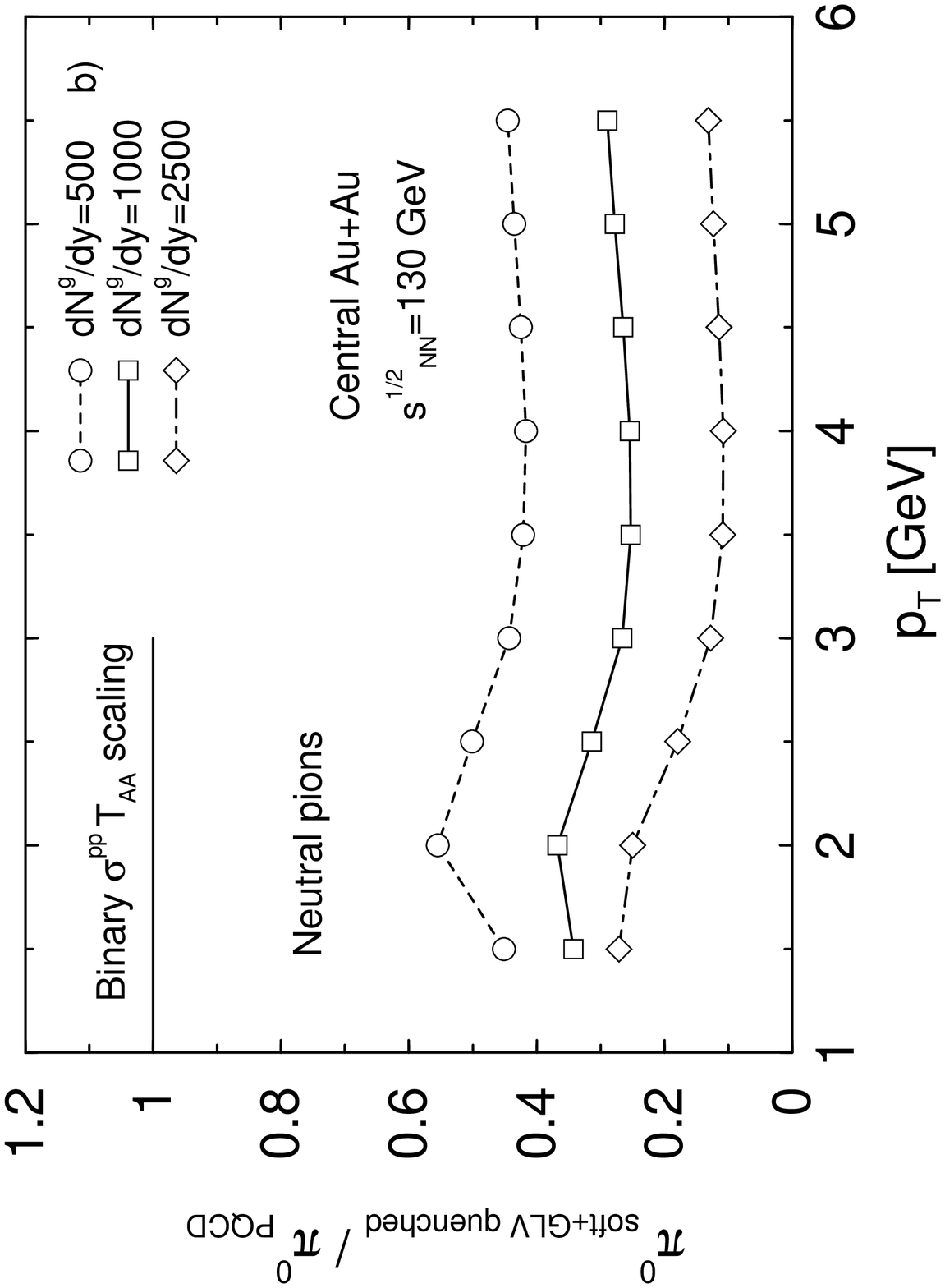}
\vspace{-2.9cm}

\vspace{2.0cm}

\begin{minipage}[t]{14cm} 
         {\small \bf Figure 2:} {\small {\bf a)}  Inclusive charged hadrons  
and  {\bf b)} neutral pions are plotted versus $p_{\rm T}$  normalized 
by the corresponding scaled PQCD multiplicities. Curves reflecting the effect of 
non-abelian energy loss driven by  $dN^g/dy = 500,\; 1000,\;2500$ are shown.
}
\end{minipage}
\end{center}
Eq.(\ref{sloperelat}) helps reduce the parameters 
that enter the phenomenological soft component  to 2: 
$T_0^\pi \sim 210 - 220$~MeV  and $dN_{ch}^-/dy \sim 600$ in central collisions
at midrapidity. Assuming that similar string and baryon junction dynamics 
drives $pp$ and $AA$ collisions, we solve for the remaining parameters  
from the the existing data~\cite{exdata} 
($\bar{p}/\pi^-\simeq 7-8\%,\;  K^-/\pi^- \simeq 12-15\%$) 
or from charge and strangeness conservation 
($\pi+/\pi^-\simeq 1,\;  K^+/K^- \simeq 1-1.1$) 
and baryon transport~\cite{junc}  ($dN^B/dy \simeq 
\beta Z \cosh (1-\alpha_B(0)y)/ \sinh (1-\alpha_B(0)Y_{\rm max}) \simeq 10-15$,
given $Y_{\rm max}=5.4,\; \beta \sim 1$).

\section{Result and conclusions}

In Figs. 2a and 2b we compute the quenching for inclusive  charged
hadrons and neutral pions. The $\sim 2$ difference in the suppression factors in
the $p_{\rm T} = 1- 4$~GeV window is largely due to the non-perturbative baryon
component of $dN_{ch}$. The comparison of the quenching factors to data~\cite{phen}
suggest initial gluon rapidity density   $dN^g/dy \sim 500 - 1000$.  
Fig.~3 presents the central result of this paper. 
The anomalous baryon/meson 
ratio exceeds unity in the $p_{\rm T}=2-5$~GeV window in central collisions. 
It is here interpreted as a combined effect of non-perturbative
baryon junction dynamics and jet quenching that suppresses
the PQCD component of pions. Fig. 3 suggests that this result may be more readily
observable in $p/\pi^+$ than in $\bar{p}/\pi^-$. The uncertainties in the mean
inverse slope of pions and protons as well as the degree of quenching  
may affect 
%the  maximum baryon excess  
$R_{B\; {\rm max}}$ by as much as
$30-50\%$ (also seen in Fig.~3), without however qualitatively changing our 
predictions.  
Boosted thermal sources calculation with $v_T=\tanh \eta_r \sim 0.6$  and
$T_f \simeq 160$~MeV
\vspace{-.1cm}
\begin{equation}
\frac{ dN_s }{dyd^2{\bf p}_{\rm T}} \sim
     m_{\rm T} \, 
K_1 \left(\frac{m_{\rm T}\cosh \eta_r}{T_f} \right)
I_0 \left(\frac{p_{\rm T}\sinh \eta_r}{T_f} \right)
\end{equation}
may produce similar low $p_{\rm T}$ behavior but $R_B$ is a
monotone function $p_{\rm T}$ and saturates at $R_B = 2$ from spin counting. 
 %In our
%two component model 
We predict that at high $p_{\rm T}>5-6$~GeV  $R_B$ decreases again
below unity to its PQCD dominated base. We also predict that in going from
central to peripheral collisions the baryon/meson anomaly disappears since
there is no jet quenching to suppress the perturbative 
pions and the $R_B$ ratio
resembles the $pp$ measurement.   Preliminary PHENIX data seems to support the 
suggested centrality dependence.

\newpage 

\begin{center}
\vspace*{7.cm}
\includegraphics{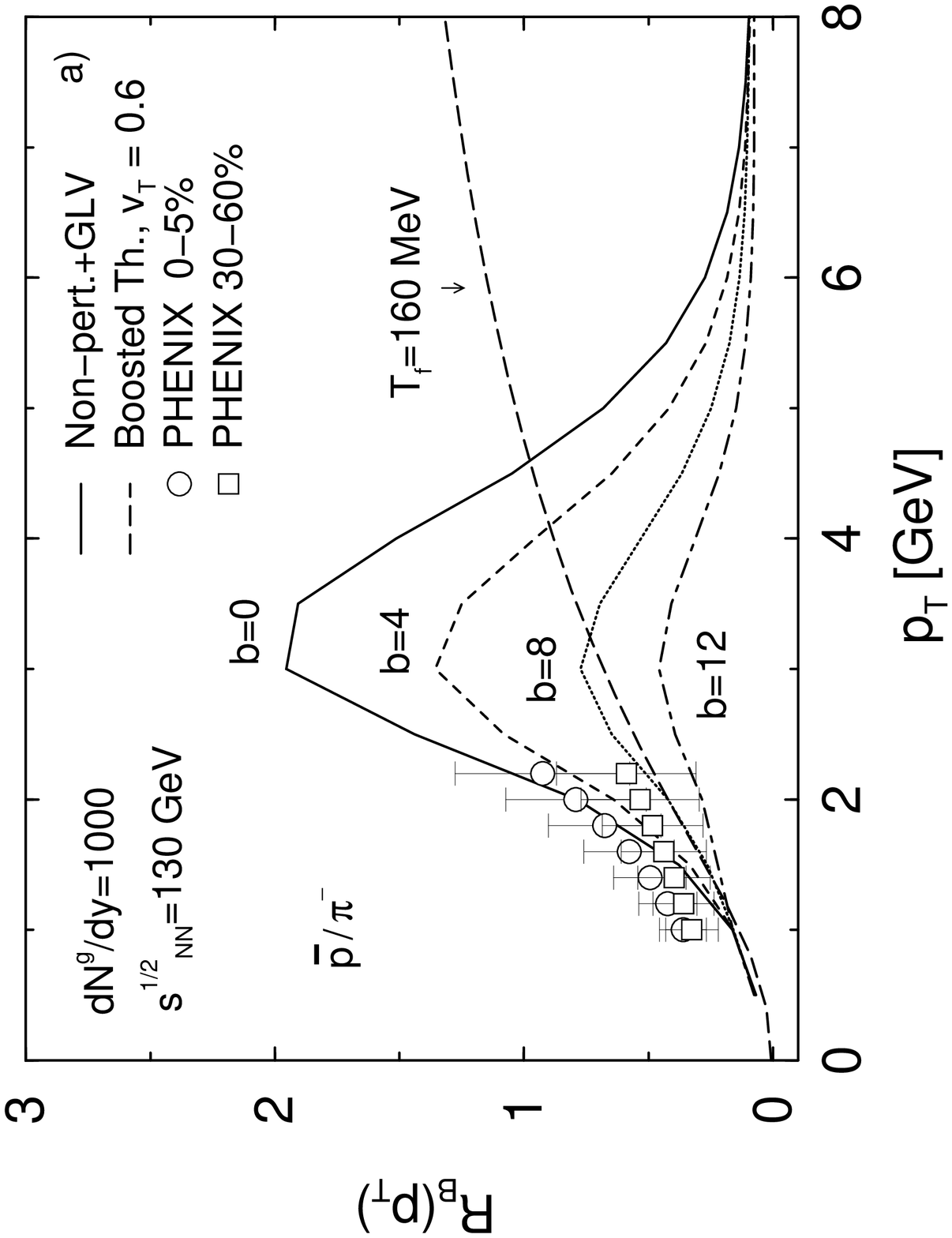}
\vspace{-2cm}
\includegraphics{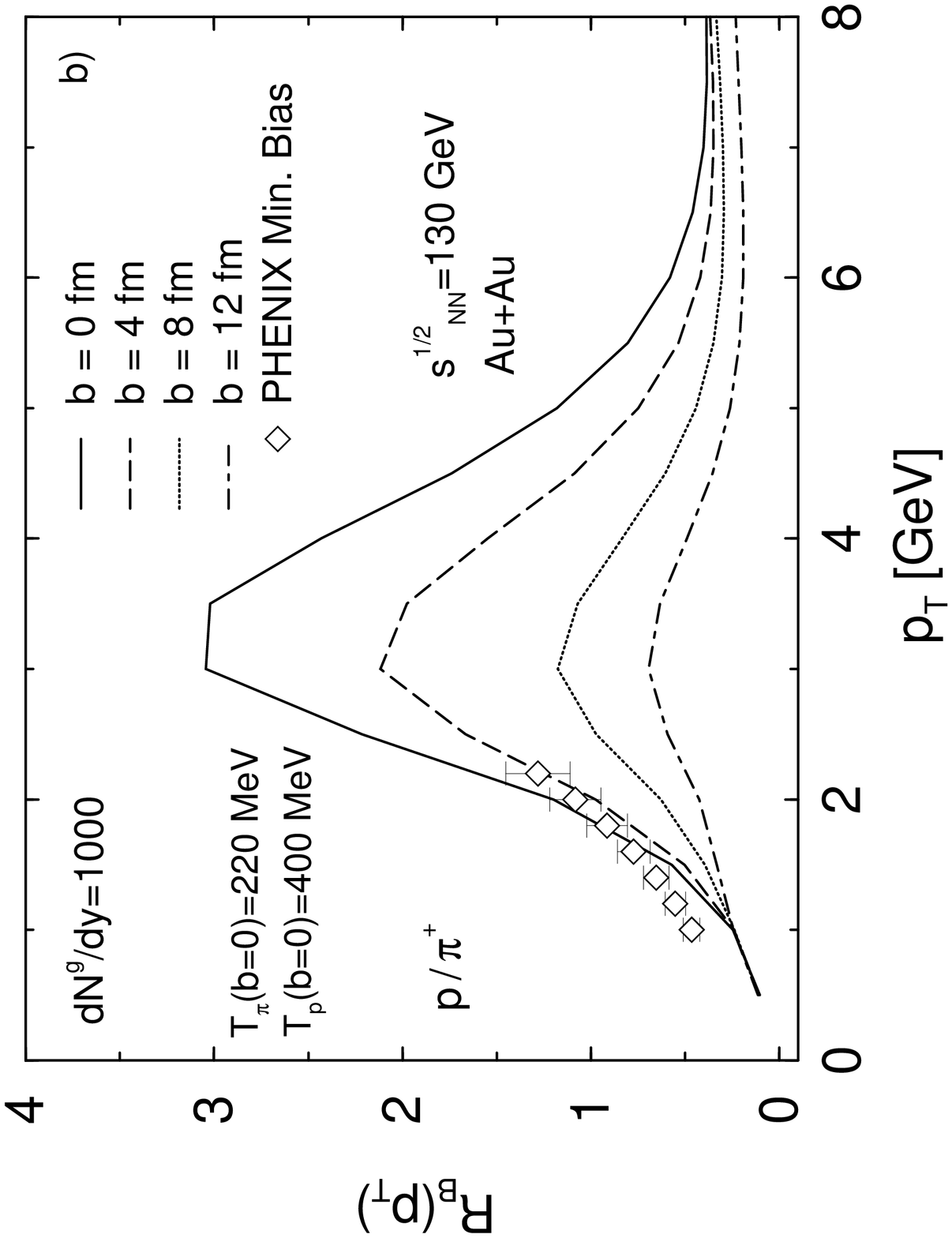}
\vspace*{7.5cm}
\includegraphics{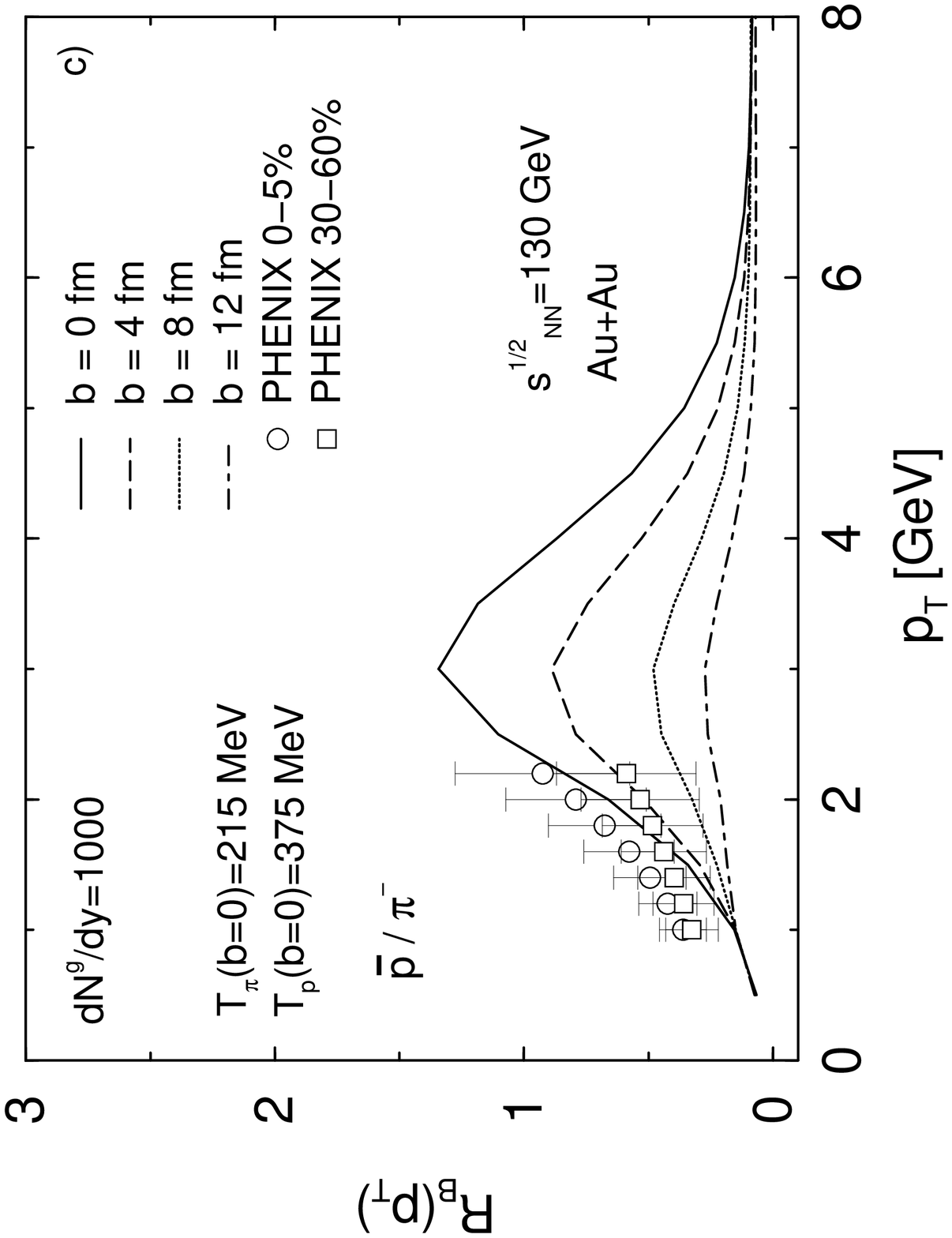}
\vspace{-2cm}
\includegraphics{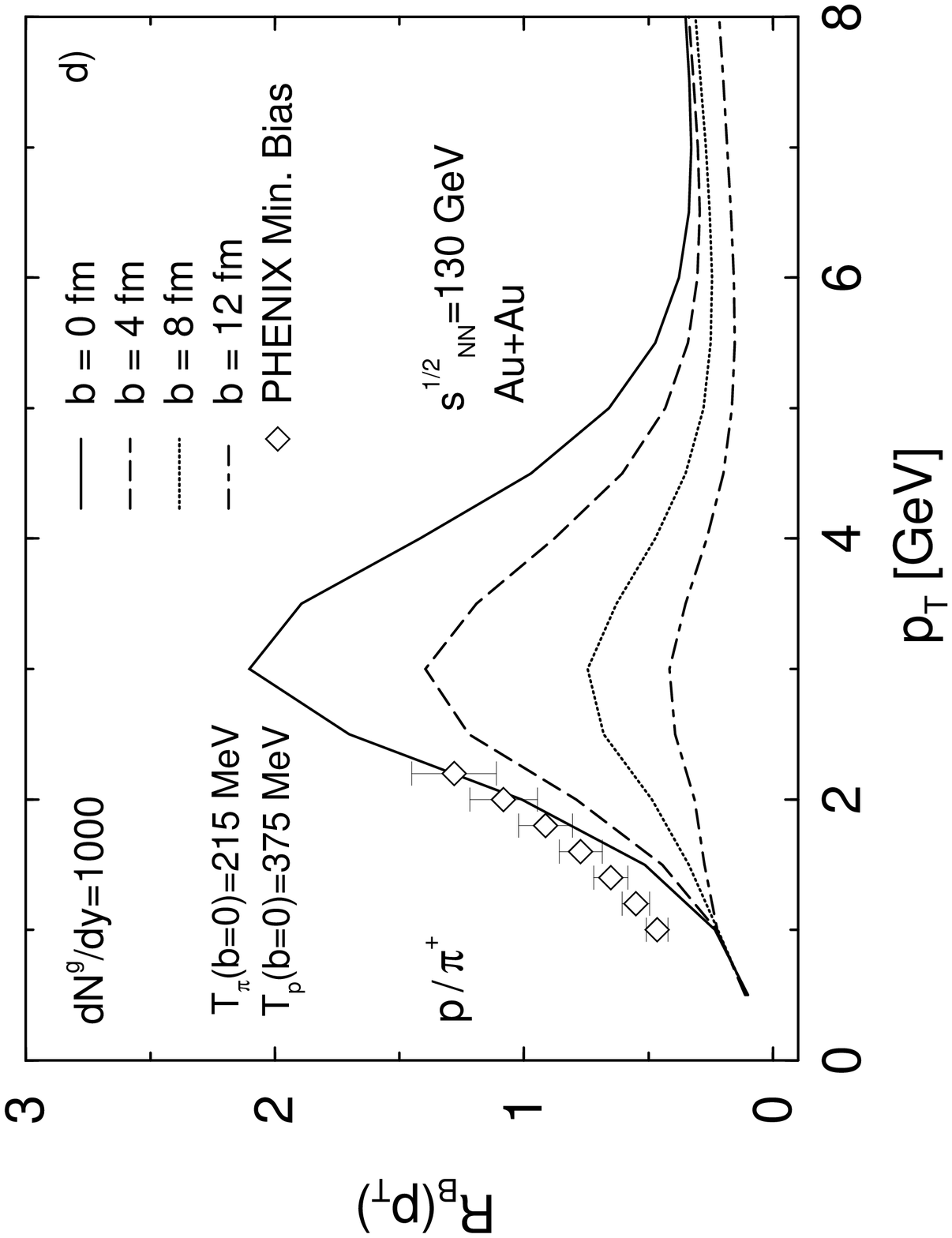}
\vspace{-2.9cm}

\vspace{2.cm}

\begin{minipage}[t]{14cm} 
         {\small \bf Figure 3:} {\small The $\bar{p}/\pi^-$ and ${p}/\pi^+$ ratios 
are plotted versus $p_{\rm T}$ for 4 different impact parameters 
${\bf b}=0,\; 4,\; 8,\; 12$~fm. 
The top and bottom row of figures illustrate
plausible uncertainties in the mean inverse slope of $\pi$ and $p$. The ratio
of {\em our fits} to preliminary PHENIX data as well as boosted thermal source 
calculation are shown for comparison.  
}
\end{minipage}
\end{center}

\end{document}